# The Fission Fragment Rocket Engine for Mars Fast Transit

John Gahl [a]*, Andrew K. Gillespie [b]*, Cuikun Lin [b], and R.V. Duncan [b]

AFFILIATIONS

[a]*Department of Electrical and Computer Engineering, University of Missouri, Columbia Missouri 65211, USA*
[b]*Department of Physics and Astronomy, Texas Tech University, Lubbock Texas 79409, USA*
*Author to whom correspondence should be addressed: gahlj@missouri.edu

Keywords: fission, fragment, nuclear, rocket, fast, transit, engine, Mars

## ABSTRACT

In this paper we discuss the advantages and challenges of utilizing Fission Fragment Rocket Engines (FFREs) to dramatically reduce transit time in space travel, for example, traveling to Mars. We discuss methods to decrease the size and weight of FFREs. These include utilizing metallic deuterides as moderators, driving the engines with electron beam bremsstrahlung, and operating the FFREs as subcritical assemblies, not as nuclear reactors. We discuss these and other new innovations based upon improved materials and technology that may be integrated into a revolutionary nuclear rocket technology.

## I. INTRODUCTION

Since the 1950's, nuclear rockets were developed primarily at Los Alamos National Laboratory to produce faster methods for space travel. These technologies utilize nuclear designs that transfer heat from a sealed core to liquid hydrogen expanders or thermionic converters in a conventional manner. Beginning in the 1980s, a more efficient nuclear energy transfer design emerged for rocketry, one which exposes the core for direct nuclear fragment thrust once the rocket is well outside the Earth's atmosphere. From FY11 through FY14, the NASA Institute for Advanced Concepts studied the Fission Fragment Rocket Engine (FFRE). The FFRE would directly convert the momentum of fission fragments into spacecraft momentum at extremely high specific impulse ($I_{SP}$). The FFRE combined the existing technologies of low-density dust trapped electrostatically (dusty plasmas) with high field magnets for fission fragment control. By designing the nuclear core to permit sufficient mean free path for the escape of fission fragments, the study showed the FFRE could convert nuclear power to thrust directly and efficiently at a delivered $I_{SP}$ of 527,000 seconds. The study showed further that, without increasing the reactor power, adding a neutral gas to the fission fragment beam significantly increased the FFRE thrust. This frictional interaction of gas and beam resulted in an engine that would continuously produce nearly 4.5 kN of thrust at a delivered impulse of 32,000 seconds, thereby reducing the (then expected) round trip mission to Mars from 3 years to 260 days. These studies concluded that the engine and spacecraft were within current technological capabilities, could be built and launched, though the FFRE engines would be very large. New materials, and recent advances in technology, promise to make FFREs much more compact and reliable.

For manned transit to Mars, speed is of the essence to minimize radiation exposure and maximize human health, as discussed in other articles in this volume. Nuclear fuel, with several million times the energy density of any chemical fuel, holds the promise of fast transit to Mars. Typically, two types of nuclear propulsion are considered for space flight: thermal propulsion and electric propulsion.

In thermal propulsion, a coolant that prevents reactor melt-down doubles as a propellant. This gas is heated by direct interaction with the fuel, which limits the gas heating to a temperature below 3000 K, where typical fission fuel melts. A metric of the efficiency of a rocket engine is the specific impulse or $I_{SP}$. The $I_{SP}$ is thrust (force) of the engine divided by the weight flow of the propellant. The higher the $I_{SP}$, the more efficient the engine. In nuclear thermal propulsion $I_{SP}$ of approximately 900 s can be achieved, compared to perhaps half that for the best chemical engines. While certainly an improvement, with nuclear energy's enormous energy density, conventional nuclear thermal propulsion could benefit from technological innovation.

In electric propulsion, the nuclear reactor is configured to produce electricity, and that electricity is used to drive a conventional ion or plasma source that provides propulsion. These sources tend to have high $I_{SP}$, but low thrust. Thrust, approximately the velocity of ejected propellant times mass flow, is required to lift a rocket out of a gravity well. For a mission to Mars, a rocket engine would ideally have the capacity to produce high thrust (for leaving the earth, moon, or mars) and then high $I_{SP}$ (high efficiency) for the interplanetary transit. As an analogy, this is similar to an automobile starting with a low gear and high torque, finally traveling at high speed in a high gear. In a reactor producing electricity, Carnot efficiency requires a large difference in output temperature compared to input temperature. In a vacuum, space craft have difficulty rejecting heat, requiring large radiators. Increasing input temperature is again limited by the melting temperature of the nuclear fuel.

For both of these technologies, nuclear electric propulsion or nuclear thermal propulsion, performance enhancement can be achieved through dealing with the fuel heating problem. If we could heat the propellant to a higher temperature than the melting point of the fuel, then we could achieve higher thrust. If we could operate the nuclear reactor at a temperature well above the melting point of the nuclear fuel, Carnot efficiency could also be improved.

## II. FISSION FRAGMENT ROCKET ENGINE

Fission fragment nuclear propulsion was proposed by George Chapline of Lawrence Livermore National Laboratory in the late 1980's.[1–3] The propulsion method depended on the fission fragments of a nuclear reaction partially escaping the fissile fuel, mitigating the heating of the fuel. When a fissile material undergoes fission, two daughter atoms are produced as fragments of the event. These atoms carry a total energy of 160 MeV and can travel at velocities up to 5% of the speed of light. If the fission occurs in a very thin layer of nuclear fuel (a few microns thick), some of the fission fragments can escape without collision, minimizing fuel heating. With a magnetic field used to direct these fragments to the exhaust direction, extremely high ISP could be achieved, perhaps on the order of $10^6$ s. This would be two orders of magnitude larger than an electric ion source and more than three orders greater than chemical or nuclear thermal propulsion. On long transit missions this increased ISP, and the decrease in mass of the rocket as its nuclear fragments are ejected, would result in an extremely high final rocket velocity. Chapline envisioned the thin fissile fuel coating being placed on carbon fibers, the fibers moving at high speed through a neutron moderator region with many other coated fibers, having sufficient fissile mass to sustain the reaction, obtaining reactor criticality. The fibers would then exit the neutron moderator to radiatively cool in the vacuum of space. Even though many fission fragments would escape the fuel and eject at high speed, the carbon fibers would still heat significantly.

To simulate a 14-MeV neutron source incident on a layer of uranium oxide, a nuclear transport code (MCNP6.2) was employed.[4] The simulation tracked any heavy ions generated to estimate the percentage that escaped as a function of the thickness of the uranium oxide layer. The results of the simulation, illustrated in **Figure 1**, confirm that nearly 100% of heavy ions can escape for layer thicknesses below a few microns. Although thicker uranium oxide layers generate more fission fragments, almost all heavy ions remain trapped inside the fissile material. Therefore, using fissile layers thicker than approximately 10 microns does not yield any benefits.

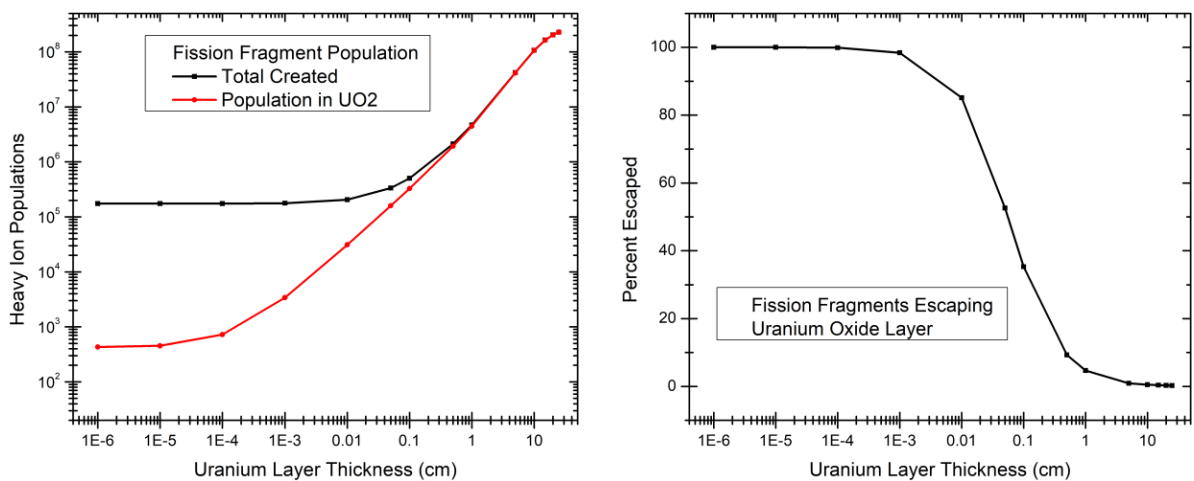

**Figure 1:** Simulation results for fission fragments in various zones around a uranium oxide layer. *Left*: The population of fission fragments either escaping or remaining inside the uranium oxide layer as a function of layer thickness. *Right*: The percentage of fission fragments escaping as a function of uranium oxide thickness.

To avoid this heating, NASA initiated the study of the Fission Fragment Rocket Engine (FFRE) concept in 2010.[5–7] The FFRE would directly convert the momentum of fission fragments into spacecraft momentum at extremely high ISP. The FFRE combined the existing technologies of low-density dust trapped electrostatically (dusty plasmas) with high field magnets for fission fragment control. The fissile fuel fragments and dust would be micron-sized, allowing for the fission fragments to escape the fuel particle without depositing significant energy into the fuel particle or heating it. By designing the nuclear core's uranium dust density to permit a sufficient mean free path for the escape of fission fragments, while simultaneously maintaining density sufficient to maintain criticality, the study showed the FFRE could convert nuclear power to thrust directly and efficiently at a delivered ISP of 527,000 seconds.

Without increasing the reactor power, adding a neutral gas to the fission fragment beam significantly increased the FFRE thrust. This frictional interaction of gas and beam resulted in an engine that would continuously produce nearly 4.5 kN of thrust at a delivered impulse of 32,000 seconds, thereby reducing the (then expected) round trip mission to Mars from 3 years to 260 days. It importantly would allow the heating of the gas propellant to a very high temperature, improving thermodynamic efficiency without melting the uranium oxide fuel. The fuel dust would still heat from the gamma radiation in the reactor, however the small size of the fuel dust would mean that the particles had a larger surface-area-to-volume ratio compared to large particles, enhancing radiative cooling of the fuel.

These studies concluded that the engine and spacecraft were within current technological capabilities, could be built and launched, however the FFREs would be very large. To sustain a nuclear reaction, a neutron produced by fission in the fuel should be moderated and available to initiate another fission event in order to "go critical." There also needs to be a sufficient amount of fuel mass to continue the reaction. In the FFRE concept, the fuel density needs to be low enough for fission fragments to get out. The design constraint is to design a reactor that keeps neutrons in, and allows fission fragments out, and fundamentally this means a low-density fuel in a large volume.

In an early conceptual design, superconducting magnets were used adjacent to the nuclear reaction. With one magnet having a radius of over 5 meters, the coils would have a combined mass of over 40 metric tons. Either a moderator or neutron reflector would be needed to keep neutrons in the reactor, maintaining criticality. This would add over 50 metric tons and the total engine mass would be over 113 metric tons.[6] In this design, the extra mass would not be expended as the rocket accelerated, substantially limiting the final velocity of the rocket.

Though there are a number of other very practical issues to consider for the successful deployment of a dusty plasma FFRE, perhaps most importantly, no one has ever built or tested a dusty plasma reactor on earth, let alone in space.

## III. INNOVATIONS ON THE FISSION FRAGMENT ROCKET ENGINE

### IIIa. Aerogel Core

Ryan Weed and others[8] have suggested a modification of the dusty plasma FFRE concept. By utilizing low density aerogel materials to contain micron-sized fuel particles, the significant complications of the dusty plasma confinement can be eliminated. The distribution of fissile particles in the aerogel can be engineered to facilitate criticality while allowing fission products to escape. The low density of the aerogel would minimize energy loss from the fission fragments while at the same time facilitating radiative cooling of the fissile particles. Spent, aged, or depleted aerogel could be simply replaced with the low mass of replacement aerogels, amounting to a trivial amount of total spacecraft mass. Recent advances in magnetic technology[9] along with reactor design versatility made possible by the utilization of aerogel could dramatically reduce FFRE mass and make deployment on "conventional" space craft, such as SpaceX's Falcon 9, possible.

**IIIb. Subcritical Assembly**

Reactors have sufficient fissile fuel and sufficient moderation and neutron reflection to sustain a nuclear chain reaction. If any of these requirements for criticality are removed, an accelerator can be used to produce neutrons to facilitate continued nuclear reactions. A reactor that cannot sustain a nuclear chain reaction is not critical and is not, by definition, a reactor. It is a subcritical nuclear assembly.

Of many examples, Gahl and Flagg[10] proposed the use of an aqueous solution of uranium salts in heavy water as a subcritical assembly. Driven by a high energy electron beam, producing x-rays upon hitting a high-Z converter, the accelerator would induce photoneutron production in the deuterium of the heavy water. The photoneutrons would induce fission in the uranium,[11] producing fission fragments for use in radio pharmacy. With more uranium and better reflection, this assembly would have been a reactor. But if kept subcritical, the device would be much easier and safer to develop, operate, and license. In a subcritical FFRE assembly, an electron accelerator could produce x-rays and subsequently photoneutrons through the irradiation of metallic deuterides. The light-element metallic deuterides would also be used as neutron moderators in the subcritical FFRE assembly, since they very effectively reduce the fission neutrons' energy through elastic collisions. A simplified example of the FFRE assembly is shown in **Figure 2**.

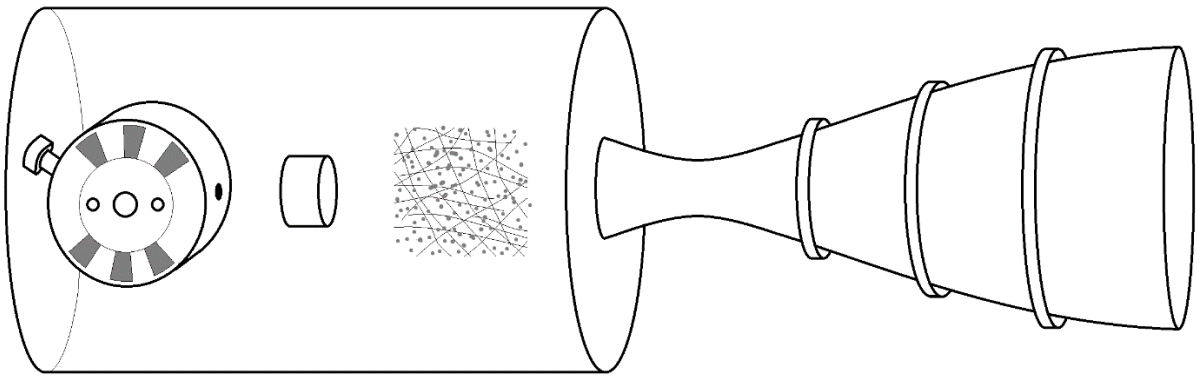

**Figure 2:** A simplified depiction of an electron beam incident on deuterated metal hydride converter. The generated bremsstrahlung x-rays create photoneutrons that drive fission in thin layers of a fissionable material.

There are many types of neutron sources available to drive fission reactions, from simple, commercially available DD fusion systems, spontaneous neutron emitting isotopes, to bremsstrahlung x-ray sources driving photoneutron reactions. Applied to FFRE technology, accelerator driven systems would be simpler to develop, much easier to control, and could facilitate the development of very small FFREs, whose geometry would never allow for a system to achieve criticality. Some of the fission fragments, such as the ones propagating on the thrust axis along the direction of travel, may be used to produce electricity, preferably by direct conversion, to power the accelerator and to meet other spacecraft needs. In addition, the small amount of waste heat generated by the subcritical assembly may be cooled through a thermoelectric converter to produce spacecraft power. **Table 1** lists neutron sources that can be used in the fission reaction. Cf-252, with a half-life of 2.65 years and $2.31 \times 10^6$ n/s per ug, is the only californium isotope that emits neutrons spontaneously. D-D and D-T reactions generate monoenergetic neutrons of 2.45 MeV and 14 MeV respectively when D+ and/or T+ ions have low kinetic energy. In recent decades, quite a few compact neutron generators using radio-frequency (RF) driven plasma discharges have been developed.[12–14] These compact neutron generators are composed of three main parts: the plasma source, the acceleration column, and the target. The source is a deuterium plasma formed by ionizing deuterium gas through RF coupling. The deuterium ions present in the plasma are then extracted from an aperture in the plasma facing electrode and accelerated up to over 100 keV onto a metal target. The target is typically either a pre-loaded metal hydride target, or a plain metal target that can adsorb deuterium and tritium atoms very efficiently and form metal hydrides. The incoming ions can then collide with the implanted deuterium through D-D or D-T fusion reactions and produce neutrons of 2.45 MeV and 14.1 MeV respectively. These neutrons can then be moderated and collimated to the necessary energies and direction by using various shielding materials such as polyethylene, heavy water, or light water.

The photoneutron can serve as another potential neutron source. Chadwick and Goldhabber discovered the nuclear photoeffect on the deuteron[15]. **Table 2** outlines the materials and their corresponding low threshold reactions, with D and Be-9 being the only nuclides possessing low binding energies, namely 2.23 MeV and 1.66 MeV, respectively. Electron accelerators can also act as neutron generators with certain advantages, including higher intensity, lower cost, and smaller size than nuclear reactors, albeit with lower flux in thermal spectra. Prior research has successfully generated neutrons using high-energy electron beams.[16–19] Tabbakh *et al* reported achieving $10^{12}$ n/cm$^2$/s, with average energies of 0.9 MeV, 0.4 MeV, and 0.9 MeV for the Pb, Ta and W targets, respectively using RHODOTRON TT200 (IBA).[20] Thus such equipment displays great potential as a neutron source. We have determined from our MCNP6.2 simulations in a typical geometry that $\sim10^4$ electrons are required to generate adequate gamma radiation to produce one neutron through the photodisintegration of a deuteron. In this configuration, each photoneutron will need to create about $10^4$ $^{235}$U fissions, so this system must be operated very close to criticality. In addition, a neutron amplifier can also be considered, comprising a cascaded series of stages consisting of fissionable fuel cells separated from each other by neutron moderator walls.[21,22] By integrating these innovations into existing designs, it may become feasible to construct FFREs with much lower masses while improving upon the ISP for reduced transit times.

**Table 1:** List of commercially available neutron and electron sources for use in driving fission reactions.

| Manufacture | | Neutrons per second | Energy (MeV) |
|---|---|---|---|
| Frontier | Cf-252 | 4.4 x 10$^9$ per Ci | 1.09 |
| Adelphi | DD-108 | 1 x 10$^8$ | 2.45 |
| | DD-109 | 1 x 10$^9$ | 2.45 |
| | DD-110 | 1 x 10$^{10}$ | 2.45 |
| | DT-110 | 1 x 10$^{10}$ | 14 |
| Starfire | nGen-100 D-D | 1 x 10$^7$ | 2.45 |
| | nGen-100 D-T | 5 x 10$^8$ | 14 |
| | nGen-300 D-D | 1 x 10$^7$ | 2.45 |
| | nGen-300 D-T | 5 x 10$^8$ | 14 |
| | nGen-400 D-D | >1 x 10$^8$ | 2.45 |
| | nGen-400 D-T | >5 x 10$^9$ | 14 |
| Thermo Fisher | P385 D-T | 3 x 10$^8$ | 14 |
| | minGen D-T | 3 x 10$^8$ | 14 |
| | minGen D-D | 6 x 10$^6$ | 2.45 |
| LLNL | Patent #6,907,097. | 10$^{12}$–10$^{14}$ | 14 |
| Electron beam accelerator [23] | | | |
| Manufacture | model | Type | Energy (MeV) |
| IBA Industrial Solutions | TT-50 | RF-SCR | 10 |
| | TT-100 | RF-SCR | 10 |
| | TT-200 | RF-SCR | 10 |
| | TT-300 | RF-SCR | 10 |
| | TT-1000 | RF-SCR | 7.5 |
| NIIEFA | UEL-10-D | RF-Linac | 10 |
| NIIEFA | Elektron 23 | DC | 1 |
| BINP | ILU-10 | RF-SCR | 5 |
| BINP | ILU-14 | RF-Linac | 10 |
| BINP | ELV-12 | DC | 1 |
| Varian | Linatron | RF-Linac | 9 |
| Mevex | Linac | RF-Linac | 3 |
| Wasik Assoc. | ICT | DC | 3 |
| Getinge Group | Linac | RF-Linac | 10 |
| Vivirad S.A. | ICT | DC | 5 |

**Table 2:** Materials and low threshold reactions.

| Nuclide | Threshold (MeV) | Maximum Cross Section Below 10 MeV [24,25] [mbarn] | reaction |
|---|---|---|---|
| $^2$D | 2.23 | 2.17 @4.8 MeV | $^2$H$(\gamma,n)^1$H |
| $^6$Li | 3.70 | 3.16 @10 MeV | $^6$Li$(\gamma,n+p)^4$He |
| $^6$Li | 5.67 | 3.16 @10 MeV | $^6$Li$(\gamma,n)^5$Li |
| $^7$Li | 7.25 | 1.09 @10 MeV | $^7$Li$(\gamma,n)^6$Li |
| $^9$Be | 1.67 | 1.41 @1.7 MeV | $^9$Be$(\gamma,n)^8$Be |
| $^{13}$C | 4.90 | 0.69 @10 MeV | $^{13}$C$(\gamma,n)^{12}$C |

## IV. CONCLUSIONS

Fast transit to Mars poses several challenges to rocket design due to overall mass, specific impulse, thrust, and thermal management. It may be possible to circumvent several of these obstacles by using subcritical assemblies, energy efficient neutron generation, and compact, high-field magnets to generate and steer fission fragments. Integrating these innovations into FFRE designs may provide increased specific impulse and controlled thrust while operating as a subcritical assembly. These new technologies and improvements may be integrated into a revolutionary nuclear rocket technology and dramatically reduce transit time to Mars.

## V. ACKNOWLEDGEMENTS

We are grateful to Matthew Looney (Texas Tech University) his suggestions and efforts as the Radiation Safety Officer involved in this effort. We thank Prof. John Brockman from MURR for his advice and recommendations associated with the radiations reported herein.

## VI. DATA AVAILABILITY

The data and methods recorded and utilized in this study can be found at https://www.depts.ttu.edu/phas/cees/, and through the Information Technology Division of Texas Tech University. Additional data that support the findings of this study are available from the corresponding author upon reasonable request.